\documentclass[twocolumn,floatfix,prl,superscriptaddress,nofootinbib]{revtex4-2}
\usepackage{graphicx}
\usepackage{amsmath}
\usepackage{url}
\usepackage{numprint}
\usepackage{hyperref}
\usepackage{braket}
\usepackage{color}

\usepackage{tikz}
\usepackage{blkarray}
\usetikzlibrary{tikzmark,decorations.pathreplacing,positioning}

\usepackage{amsfonts}
\usepackage[makeroom]{cancel}

\graphicspath{{Graphs/}}

\newcommand{\cnot}{\mbox{\small CNOT}}

\begin{document}

\title{Imperfect-Information Games on Quantum Computers: A Case Study in Skat}
\author{U.~Armbr\"uster}\normalfont
\affiliation{Atos Future Makers Research Community, Wagramer Straße 19, 1220 Vienna}
\author{S.~Edelkamp}\normalfont
\affiliation{Computer Science Department, Faculty of Electrical Engineering, Czech Technical University, Prague}
\affiliation{Department of Theoretical Computer Science and Mathematical Logic
Faculty of Mathematics and Physics, Charles University in Prague}
\author{G.~Maresch}\normalfont
\affiliation{Institute for Discrete Mathematics and Geometry, Technical University of Vienna, Wiedner Hauptstraße 8-10, Vienna}
\author{E.~Schulze}\normalfont
\affiliation{Bull GmbH, Von-Der-Wettern-Straße 27, 51149 Cologne}

\begin{abstract}

For decades, it has been known that quantum computers might serve as tools for solving specific classes of problems that have long been considered computationally infeasible. Some of these problems are combinatorial, with the quantum advantage arising from the rapidly growing size of the corresponding decision tree. Even though the winning move in a specific imperfect-information situation cannot generally be determined, the relevant question is which choice yields the best long-term outcome. This leads to the search problem of maximizing the payoff function. This paper illustrates how quantum computers can play a significant role in solving such games, using the popular German card game Skat as an example. To this end, quantum registers are used to encode the game's information, and corresponding quantum gates are constructed to model the progress of the game while respecting its rules. Finally, a score operator is used to project the quantum state onto the winning subspace and to evaluate the winning probabilities using quantum algorithms, such as quantum counting of winning paths, to obtain a potential computational speedup over classical approaches. This yields a reasonable recommendation for the next action at the table. This approach is computationally infeasible on a classical computer because of the size of the tree-search problem. Peculiarities of the problem are discussed that may lead to a quantum advantage beyond a certain problem size.

\end{abstract}
\maketitle

\section{Introduction}

Game theory has had a major impact on many sectors beyond games. A well-known example of a game with imperfect information, in which players have only partial knowledge of the current game state and its variables, is an economic model of labor-management negotiation. Some of the predicted economic variables are only partially known through probability distributions. Therefore, despite improvements in models for predicting economic variables, it is desirable that uncertainty leads to an equilibrium state of decisions corresponding to the formal game-theoretic solution of the system. While it remains to be determined whether the concrete problem is too complex for a classical solution, several more general aspects of imperfect-information games should be considered in advance.

More specifically, it has been shown that $n$-player imperfect-information games may be NP-hard~\cite{Blair}, making them difficult to compute classically. To solve them formally, or at least approximate the game-theoretic solution, significant advantages may be obtained from the computational power of entangled quantum systems.

A well-known statement by the mathematician and computer scientist Hans-Joachim Bremermann from the 1960s reads~\cite{Bremermann}.

\textit{"In order to have a computer play a perfect or nearly perfect game, it will be necessary either to analyze the game completely ... or to analyze the game in an approximate way and combine this with a limited amount of tree searching. ... However, a theoretical understanding of such heuristic programming is still very much wanting."}

Since then, game theory has developed in many directions, including the use of artificial intelligence and machine learning as well as other emerging technologies. At that time, when computers were not yet able to beat humans in most complex strategic and combinatorial games, quantum computers were expected to remain only a prospect for the foreseeable future. This has changed drastically in recent years with new theoretical concepts and hardware implementations of real quantum computers.

Generally, a gate-based quantum computer is a device that manipulates quantum states unitarily. Today, quantum computers are considered capable of solving problems that, until recently, were thought to be intractable~\cite{Shor_1997}. 
Some complex combinatorial problems are equivalent to, or can be mapped to, problems in mathematical game theory. Game theory can be used to model decision processes, raising the question of which individual strategies a player or party should follow to optimize the outcome of a predefined target function. Since some games are extremely difficult to solve completely, new computational methods may be necessary to treat them as large optimization problems.

Some of the most complex games, such as chess or Go, were considered computationally infeasible for decades. The reason is the enormous computation time, estimated by Shannon in 1950 as approximately $\sim 10^{90}$ years on a 2 kHz computer for around $10^{120}$ possibilities. 

In terms of state-space size alone, a quantum computer with roughly 400 logical, error-corrected qubits~\cite{Dalzell_2020} might be able to handle it. However, many board games, such as checkers~\cite{checkers} and Nine Men's Morris~\cite{ninemenmorris}, have already been solved on traditional hardware. In others, such as chess, Shogi, and Go, computers clearly outperform humans~\cite{alphazero}.

Card-game play has recently become a research objective for decision making under imperfect information,
and current card-game solvers are beginning to challenge human pre-eminence. 
Notably, there is research on Bridge~\cite{DBLP:journals/corr/abs-1911-07960} and
Skat~\cite{DBLP:conf/socs/Edelkamp19}.
This work shows that quantum computing is suited to deal with large state spaces in incomplete-information games, using Skat as a case study.

As a card-game, Skat~\cite{Schettler} poses many combinatorial search challenges, such as cooperative and adversarial behavior among the players, randomness in the deal, and partial knowledge due to hidden cards. With three players and 32 cards, it is more compact than Bridge, but the additional uncertainty introduced by the two cards in the Skat creates many subtleties. 

\section{Quantum Game Theory}

Around 25 years ago, quantum game theory began to attract attention~\cite{Meyer_1999,Eisert_1999}. The idea was to use entangled initial states and quantum correlations to establish cooperation that is not possible classically. 

This approach was applied to simple, well-known problems in game theory, such as the Prisoner's Dilemma and the Mean King's Problem. As a result, quantum strategies can lead to new kinds of equilibrium states.

The concept of a classical player, corresponding to a simple random variable on a finite-cardinality probability space, has been shown to extend to that of a quantum player, corresponding to a self-adjoint operator on a quantum-probability Hilbert space. In this setting, quantum versions of von Neumann's minimax theorem for zero-sum games---often considered the starting point of game theory---were proven in~\cite{quantumminmax}.

For clarity, it should be emphasized at the outset that there is a fundamental difference between a quantum solver for a classical game and a genuine quantum game, in which a quantum version of a game is defined by new rules that often have very little to do with the original game.

\section{Solving Imperfect-Information Games}

Rather than discussing these quantum versions of classical games in greater detail, the purpose of this paper should be stressed. The purpose is to use the power of quantum computation to \textit{solve} the classical game. Therefore, we must clarify what constitutes a solution to a game of this kind. Thus, classical game theory is considered first.

Today, imperfect-information games are solved using different methods. Finding Nash equilibria via the Lemke--Howson algorithm~\cite{Lemke}, modeling the problem as a Bayesian game in which uncertainty is represented by a probability distribution~\cite{Harsanyi}, and applying so-called counterfactual regret minimization~\cite{Zinkevich}, in which the regret function of alternative strategies is minimized, are among the most established methods for finding the best strategies for games of this category. However, deep-learning algorithms and Monte Carlo tree search have recently become increasingly popular for more complex games.

Which of these methods is most suitable depends strongly on the problem itself, the available resources, and the desired precision of the solution. Thus, this point cannot be clarified in general without considering the game itself in more detail.

\section{Skat in Classical Game Theory}

Skat~\cite{Schettler} is a fascinating imperfect-information three-player
 card game\footnote{For the rules of Skat, see \url{www.pagat.com/schafk/skat.html}.}. 

The game was invented in the middle of the 19th century in Altenburg, Germany. Although Skat has been played widely for more than a century, only a few authors have systematically investigated its underlying theory and game strategies. An early approach was given by Emanuel Lasker, the first and only German chess world champion~\cite{Lasker,Lasker2}, who suggested basic rules primarily intended to help beginners address the problems encountered in the game. 

\begin{figure}[t]
    \centering
    \includegraphics[width=9cm]{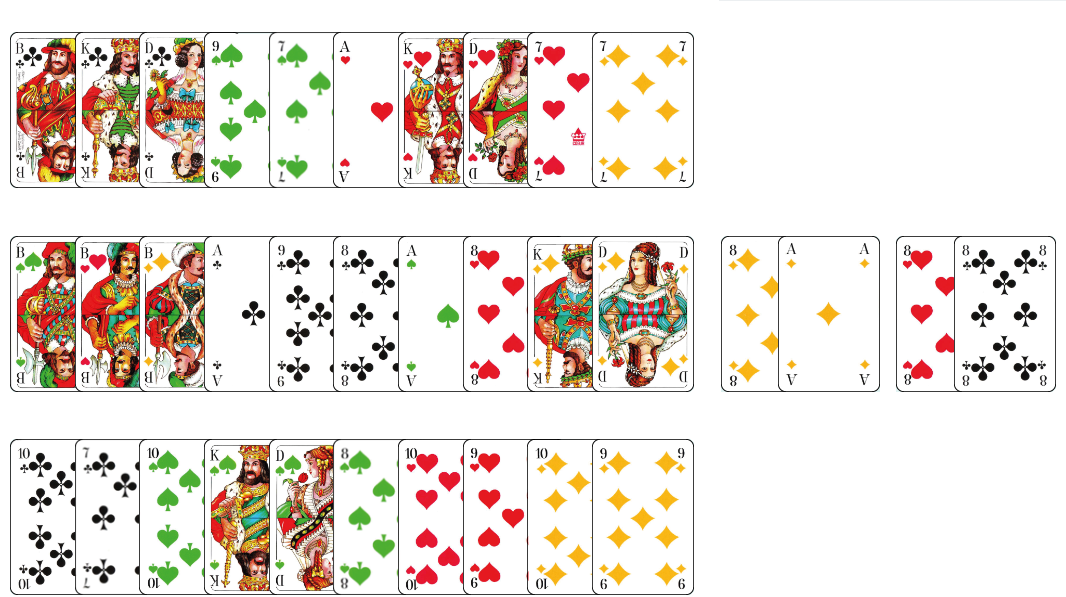}
    \caption{Skat deal with the Diamond 8 and Ace taken, and the Heart and Club 8 discarded by the declarer in the middle.}
    \label{fig:skatput}
\end{figure}

The game of Skat has been studied in many books~\cite{Wergin,Grandmontagne,Kinback,Quambush,Harmel,Rainer}. It shares similarities with Marias(ch), played in the Czech Republic and Slovakia, and Ulti, played in Hungary. At the beginning of a game (see Fig.~\ref{fig:skatput}), each player receives ten cards, which are hidden from the other players. The remaining two cards, called the Skat, are placed face down on the table. 

After the cards have been dealt, the game has four stages:

\begin{description}
\item 
[bidding] where the players try to become declarer by announcing lower bounds for the contract; 
\item 
[game selection] after taking the Skat, the winner of the bidding decides which game is to be played; 
\item 
[skat putting] the declarer discards two cards, if the Skat was taken, and announces the game to be played;  
\item 
[trick-taking] in clockwise order, each player places a card on the table, and the winner of the trick leads the next trick.
\end{description}

Research work in Computer Skat includes~\cite{skatlong,DBLP:journals/corr/abs-1905-10911,DBLP:conf/cg/Edelkamp22,DBLP:journals/corr/abs-1905-10907,DBLP:conf/socs/Edelkamp19,ownskat,ownelo}.
The \emph{double-dummy Skat solver}, a fast open-card Skat solver~\cite{diplomkupfersc}, was extended to partially observable gameplay using Monte Carlo sampling. The lack of information exchange between the players, however, has led to efforts to apply machine-learning methods~\cite{biddingskat,DBLP:conf/ijcai/BuroLFS09,DBLP:journals/corr/abs-1903-09604,DBLP:journals/corr/abs-1905-10907,DBLP:journals/corr/abs-1905-10911,skatfurtak}. Later on, knowledge-based paranoia search~\cite{DBLP:conf/cig/Edelkamp21} and a weighted sampling of the belief space were used~\cite{DBLP:journals/corr/abs-1905-10907,DBLP:conf/ecai/CohensiusMOS20}.
Together with the inclusion of hope cards,
this made it possible to beat a top player once in an unofficial match~\cite{DBLP:conf/cg/Edelkamp22}.

\begin{figure}[t]
\centering
\includegraphics[width=0.49\textwidth%,natwidth=510,natheight=542
]{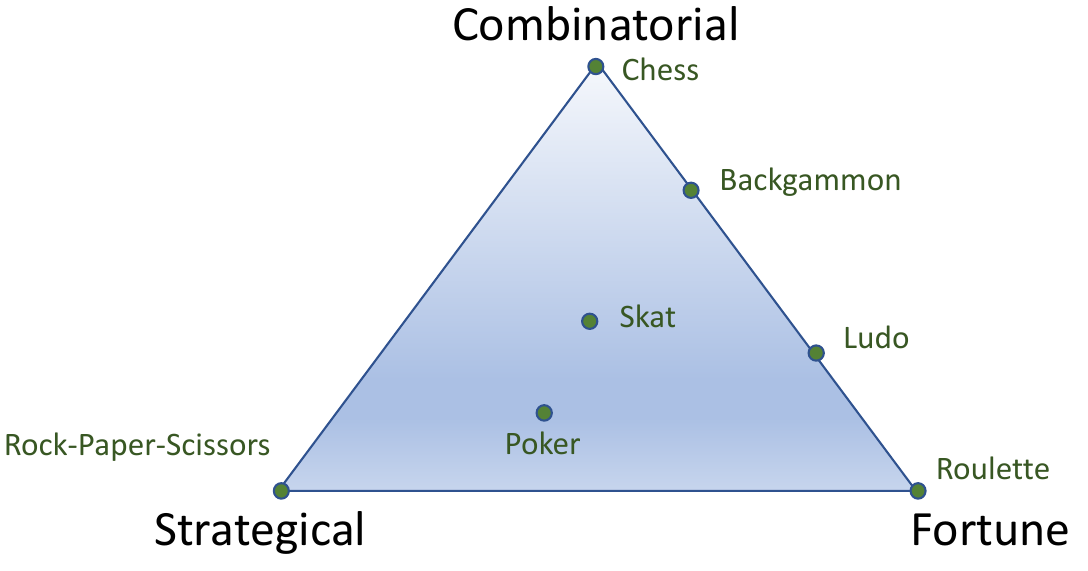}
\caption{Skat, classified among other games in terms of strategic, combinatorial, and fortune aspects in the Bewersdorff triangle.}
\label{graphic_games}
\end{figure}

To make Skat comparable to other games and to previous efforts to solve them, Skat is classified accordingly. Fig.~\ref{graphic_games} provides a rough characterization of games in terms of combinatorial, strategic, and fortune aspects. 
While chess and other purely combinatorial board games are complex mainly because of the \textit{de facto} unlimited number of moves, the difficulty in certain card games has a different origin, namely the lack of perfect information.  
The game is only partially observable, and reasoning has to be performed in the so-called belief space, the set of all possible worlds. Besides sampling, one option is $\alpha\mu$ search~\cite{DBLP:conf/ijcai/LiZCV22},
including some of its refinements~\cite{isaim},
or $\alpha\beta$ search for {AND-OR} graphs with partially ordered values~\cite{DBLP:conf/ijcai/LiZCV22}.

To determine what kind of problem has to be solved, it is first necessary to clarify what exactly is meant by a \textit{solution} of the game and what precisely has to be maximized. Disregarding some peculiarities such as the bidding process and Skat putting, for simplicity, a game consists of 30 cards being played in a certain order. The number of permutations of these 30 cards gives an upper bound for the possible games that have to be computed, but it is greatly reduced by the rules, which disallow many of those permutations. A solution of the game would be a card permutation that is allowed by the rules and leads to the highest value of the payoff function, while being in a stable equilibrium in which all players are satisfied with their respective choices according to the expectation value.

The major limiting factor in computing such a solution is therefore the tree search. More precisely, the number of winning paths for all possible distributions leads to a large combinatorial problem. According to the estimates below, computing all these paths classically would take up to 8.7 million years. Therefore, new computing technologies are suggested to tackle this part separately.

\section{Towards Quantum Skat}

While all relevant information is visible on a chessboard, an opponent's cards remain hidden. Thus, educated guesses are required to maintain a reasonable basis for decisions in the game. In fact, the game can be regarded as being in a superposition of all possible distributions at the beginning, which is reduced as the game proceeds.

Unlike quantum chess and tic-taq-toe, the rules of the game do not have to be changed; quantum computing is used to analyze incomplete-information games.
Although Skat is studied here, the results should be transferable to other card games and games with incomplete information. 
For example, Stratego is a classical two-player zero-sum game
with partial observability. With DeepNash~\cite{Perolat_2022}, an AI was able to learn the board game from scratch up to human expert level. 
Without discussing the rules and gameplay of Skat in detail, the raw numbers already provide an indication of quantum feasibility.

In quantum mechanics, the general state of a qubit can be represented as a linear superposition of its two orthonormal basis states (or basis vectors), denoted by $|0\rangle ={\bigl [}{\begin{smallmatrix}1\\0\end{smallmatrix}}{\bigr ]}$ and $|1\rangle ={\bigl [}{\begin{smallmatrix}0\\1\end{smallmatrix}}{\bigr ]}$
in Dirac notation. These two orthonormal basis states span the two-dimensional Hilbert space of the qubit.

% A set of qubits taken together is called a quantum register. For example, two qubits could be represented in a four-dimensional linear vector space spanned by the following product basis states:
% $${\displaystyle |00\rangle ={\biggl [}{\begin{smallmatrix}1\\0\\0\\0\end{smallmatrix}}{\biggr ]}}, 
% {\displaystyle |01\rangle ={\biggl [}{\begin{smallmatrix}0\\1\\0\\0\end{smallmatrix}}{\biggr ]}}, {\displaystyle |10\rangle ={\biggl [}{\begin{smallmatrix}0\\0\\1\\0\end{smallmatrix}}{\biggr ]}}, \mbox{and}\ {\displaystyle |11\rangle ={\biggl [}{\begin{smallmatrix}0\\0\\0\\1\end{smallmatrix}}{\biggr ]}}.$$

In general, $n$ qubits are represented by a state vector in a $2^n$-dimensional Hilbert space. For the initial state $\Phi_{\rm ini}$ in a card game, this gives 
\begin{align}
\ket{\Phi_{\rm ini}}=\frac{1}{\sqrt{\mathcal{N}}}\cdot\left(\ket{\Phi_1}+\ket{\Phi_2}+ \dots + \ket{\Phi_\mathcal{N}}\right).
\label{eq:initial}
\end{align}
where each $\Phi_i$ represents one possible card distribution, and $\mathcal{N}$ is the number of possibilities, which is yet to be determined and serves as a normalization constant.

A game of Skat starts with three players, each receiving ten cards from a 32-card deck (7--8--9--10--J(ack)--Q(ueen)--K(ing)--A(ce), in four suits: Diamonds, Hearts, Spades, and Clubs). The two remaining cards are not distributed to the players in the initial deal, but remain accessible through a bidding process that any player can win to become soloist and declarer.

After sketching the main idea and introducing the fundamental game mechanics, the focus now shifts to the implementation of the problem in a corresponding quantum algorithm. Therefore, a quantum-mechanical description and encoding of the problem in terms of quantum states is needed first, so that a gate set can be defined that properly represents the progress of the game.

\section{Encoding}
In principle, there are many different ways to map a given card distribution to a state vector. Moreover, the cards may serve different purposes depending on the evolution of the game. While it remains to be seen whether the following approach is practical, it is certainly possible to fix a specific permutation of the 32 cards once and for all (e.g., in a non-increasing order according to the trick-taking partial order, cf. \autoref{fig:tricktakingpo}). Thus our state-vector takes the form
\begin{align}
\ket{\Psi}&= \bigotimes_{i=1}^{32} \ket{\psi_i},
\end{align}
where each $\ket{\psi_i}$ contains all the relevant information in the game about the card with positional index $i$. To capture this information we construct
these kets as
\begin{align}
  \ket{\psi_i} = \ket{\psi_i^{(1)}}\ket{\psi_i^{(2)}}\ldots\ket{\psi_i^{(n_i)}},
\end{align}
where each $\ket{\psi_i^{(j)}}$ is a 1-qubit state. We choose $n_i$, the number of \emph{card-qubits}, to be constant across all cards and denote this number by $n_c$. Thus, each $\ket{\psi_i}$ is an element of a $2^{n_c}$-dimensional Hilbert space $\mathcal{H}_c$, and the full initial state $\ket{\Psi}$ is an element of the $2^{32 n_c}$ dimensional Hilbert space $\mathcal{H} = \left(\mathcal{H}_c\right)^{\otimes 32}$.

We refer to the basis states of $\mathcal{H}_c$ in the computational basis as \emph{card states}. Each card state can be represented as an $n_c$-digit binary number. 
This encoding of card states will prove crucial for keeping track of the evolution of the game.
In principle, one could simply list all states a card could possibly be in and assign binary numbers to these states, but more suitable encodings are likely available. For the game of Skat, the card state can be decomposed as 
\begin{equation}
\ket{q_{j_0}\ldots q_{j_{n_c-1}}} = \ket{\rm player}\ket{\rm action}\ket{\rm auxiliary}.
\end{equation}

\begin{table}[h]
    \centering
    \caption{Encoding the card state $\ket{q_{j_0}q_{j_1}}\ket{q_{j_2}q_{j_3}q_{j_4}}$, auxiliaries omitted.}
\begin{tabular}{|r|c|c|}\hline
        positions & encoded state  \\ \hline
         $j_0j_1$ &  fore-, middle-, rearhand, skat\\
         $j_2j_3j_4$ & card location (hand, table, stack, skat) \\
         %$j_5$ & auxiliary\\ 
         \hline
\end{tabular}
\label{tab:encoding}
\end{table}
When this encoding is used, the $q_{j_2}$ qubits will occasionally be renamed as $q_t$ and referred to as \emph{table}-qubits, and the $q_{j_3}$ qubits as \emph{stack}-qubits $q_s$. The state of the qubit $q_{j_4}$ will determine whether suit has to be followed. All auxiliary qubits are collected in $\ket{\rm auxiliary}$. More than one auxiliary qubit may be needed %, cf. \autoref{subsec:tricktaking}
and the labeling of these qubits will be specified where needed.

A typical card first stays in the hand of a player, $\ket{\alpha}\ket{000}$, then is played and placed on the table as part of a specific sequence, $\ket{\alpha}\ket{01q_{j_4}}$, and finally ends up in the stack of a possibly different player, $\ket{\alpha'}\ket{11q_{j_4}}$. 

It is important to note that the details of the card-state encoding determine the structure of an actual quantum-circuit implementation, but do not matter from an abstract point of view. 

Moreover, this encoding is not optimized, and its redundancies are obvious. Nevertheless, it is easy to work with and will therefore be used for now. For Skat we expect $n_c\le 5+3=8$, so a total of $8\cdot 32= 256$ qubits are needed. A more efficient card-state encoding could possibly reduce this to $5\cdot 32=160$ qubits or even less.

 \subsection{Game Progress and Quantum Gates}

Quantum gates are used to manipulate quantum states.
There are different types of quantum gates, such as single-qubit gates, which can, for example, flip the state of a qubit or allow superposition states to be created.
There are also two-qubit gates. These allow qubits to interact and can be used to create quantum entanglement, a state of two or more qubits correlated in a way that cannot be obtained using classical bits. In contrast to classical gates, all quantum gates are unitary, which means, in particular, that they are reversible.

It is a non-trivial task to single out all state vectors $\ket{\phi_i}\in \mathcal{H}$ that correspond to a valid initial card distribution in the game of Skat. In particular, $\ket{\Phi_{\rm ini}}$ will \emph{not} be a superposition of all basis states, but only of a relatively small subset. In the examples, the approach presented in~\cite{superposition} is followed, and a deterministic Boolean function $f_{\rm valid}\hspace{-3pt}: 2^{{\rm dim}\,\,\mathcal{H}}\to \{0,1\}$ is constructed to prepare the initial state \eqref{eq:initial}. $f_{\rm valid}$ maps a distribution to 1 if it corresponds to a valid card distribution, otherwise to 0. The corresponding unitary operator, which maps $\ket{0 \ldots 0}$ to $\ket{\Phi_{\rm ini}}$, is denoted by $U_{\rm ini}$.

On the qubit level, each step in the evolution of the game is effected by a unitary operator acting on the state vector. The rules of the game determine the structure of these unitaries.  
From a practical point of view, it is desirable to decompose the unitaries into parts that act only on one card, i.e.,
\begin{align}
    U = \bigotimes_{i=1}^{32} U_i,
    \label{eq:tensorunitary}
\end{align}
where each $U_i$ acts on $\mathcal{H}_c^{(i)}$, but $U_i = I$ for all but one positional index. Furthermore, it will prove useful to represent each $U_i$ as a sum of Pauli operators
\begin{align}
    U_i = \sum_k c_{k,i} \cdot \Pi_0^{(k)}\otimes \ldots \otimes \Pi_{n_c-1}^{(k)},
    \label{eq:paulirepresentation}
\end{align}
where $\Pi_j^{(k)}\in \{I, X, Y, Z\}$. Unfortunately, the relatively simple structure of \autoref{eq:tensorunitary} is possible only for some phases of the game, while for others, a more intricate structure arises.

\subsection{Playing one card}
A basic operation in any card game is playing a card, i.e., transferring a card from a player's hand to the table. There is a \emph{caveat}: a card can only be placed on the table if it is currently in a player's hand. This means that the transformation of the card state has to be conditioned on the card being in a player's hand.

With the encoding proposed in \autoref{tab:encoding}, this amounts to changing the card state\footnote{Omitting the inactive suit-following qubit}\, $\ket{\rm action} = \ket{00}$ to $\ket{\rm action}=\ket{01}$, which corresponds to applying a single Pauli-X gate, subject to the condition that the table qubit is in state $\ket{0}$. Here, a problem arises: a qubit cannot be both the control \emph{and} the target of a controlled operation. Thus, one \textcolor{blue}{auxiliary} qubit per card has to be used. The corresponding unitary must transform $\ket{0\textcolor{olive}{0}}\ket{\textcolor{blue}{0}}$ to $\ket{0\textcolor{olive}{1}}\ket{\textcolor{blue}{0}}$.
The basic quantum circuit $C$ for this unitary involves only the $\textcolor{olive}{q_t}$ and $\textcolor{blue}{q_a}$ qubits:
\begin{equation}\label{eq:play_card}
    C=\textcolor{olive}{I}\otimes \textcolor{blue}{X} \cdot \cnot  \cdot \textcolor{olive}{I}\otimes \textcolor{blue}{X},
\end{equation}
where $\textcolor{olive}{q_t}$ is the target and the auxiliary  $\textcolor{blue}{q_a}$ is the control.

\subsection{Playing all cards simultaneously}
At each player's turn, quantum capabilities are used to represent all playable cards simultaneously, i.e., to prepare a superposition
\begin{align}
    \ket{\Phi} = \frac{1}{\sqrt\mathcal{N}} \left( \ket{\Phi_{i_1}} + \ket{\Phi_{i_2}} +\ldots +\ket{\Phi_{i_{\mathcal{N}}}} \right),
\end{align}
where each $\ket{\Phi_{i_j}}$ corresponds to the situation in which the $j$-th card allowed by the rules has been played. A first approach to implementing this is to construct, for each $N$-card subset $\mathcal{S}$ of the 32-card deck a unitary operator $U_{\mathcal{S}}$ controlled by the $N$ auxiliary qubits $q_a$ 
and targeting the $q_t$ qubits of each card (using the notation from the paragraph above). While this may be feasible if only a few cards remain to be played, it is certainly not feasible when more cards remain, e.g., with five cards left in hand, there are still ${32 \choose 5} \approx \numprint{200000}$ possibilities).
Since these difficulties are serious for the implementation of the full game, more modest settings are considered in the examples. These settings nevertheless already exhibit some characteristics of the 32-card case.

\subsubsection{Construction of the controlled $CP^n_k$ gate}
The \emph{controlled card-play} gate is introduced, which facilitates playing one of $k$ cards from the hand out of a deck of $n$ cards. Its uncontrolled version is a $k$-qubit gate that maps $\ket{0\ldots 0}$ to the equally weighted superposition
\begin{equation}\label{eq:cp_action}
\frac{1}{\sqrt{k}}\left(\ket{10\ldots 0}+\ket{01 \ldots 0}+\ket{0\ldots 01} \right).
\end{equation}

In the following, the gate is constructed for the action of forehand player A, whose ID is assumed to be encoded in the player-qubit state $\ket{\rm player} = \ket{00}$. This is only to simplify notation, since versions of this gate for other players are constructed analogously. Note that this construction does not yet take suit-following into account, and therefore does not rely on the corresponding qubit from \autoref{tab:encoding}.

For $k=2$, the unitary gate that implements \autoref{eq:cp_action} can be expressed using the well-known Hadamard gate: $\cnot \cdot H\otimes I.$ For $k>2$, the gate can be constructed using the algorithm proposed in~\cite{superposition} via the Boolean function
\begin{align}
    &f:\, \{0,1\}^k \to \{0,1\}\\ \notag
    &f(\vec x) = 1, \textrm{if } x_i = 1 \textrm{ for exactly one } i.
\end{align}
This gate will be denoted by $SP_{i_1,\ldots, i_k}$. 
Note that this gate acts on the table qubits $\ket{q_t^{(i)}}$ of the $k$ involved cards.   To extend $SP_{i_1,\ldots, i_k}$ to a unitary operator on $\mathcal H$, it is defined to coincide with the identity operator on the orthogonal complement of the involved qubits.

%Here $q_{\vec a}$ denotes the involved auxiliary qubits (one for each card).
For each card index $i$, let $X_i$ be the Pauli-X operator that flips the $q_t$ qubit of card $i$. In the next step, a $k$-tuple $i_1<\ldots< i_k\le n$, abbreviated by $\vec i$, is fixed, and conditions are imposed on the operator 
\begin{align}\label{eq:uncontrolled_cardplay}
    C_{\vec i}:=X_{i_1}\cdots X_{i_k}\cdot SP_{i_1,\ldots, i_k}
\end{align} which acts on the $\ket{q_t}$ qubits of each of the $k$ card states.

To ensure that the operator $C_{\vec i}$ is applied only to cards that are allowed to be played, the tuple $q_{\vec p}=(q_{0}^{(1)},q_{1}^{(1)};\ldots;q_{0}^{(n)},q_{1}^{(n)})$ of player qubits and the tuple $q_{\vec a}$ of auxiliary qubits are used to control this gate: the operator in \autoref{eq:uncontrolled_cardplay} is conditioned on those tuples in which exactly $k$ of the $n$ kets $\ket{q_{0}^{(l)} q_{1}^{(l)}}\ket{q_a^{(l)}}$ are in state $\textcolor{blue}{\ket{00}}\ket{0}$, corresponding to $k$ cards in the hand of player \textcolor{blue}{A}. Denoting the set of all ordered $k$-element subsets of the index set by $\mathcal{C}_k$, the resulting unitary can be written as
\begin{align} \label{eq:card_playing_gate}
CP_k^n=    \prod_{\vec i \in \mathcal{C}_k }C_{\,\vec i}\,\big| (\vec \alpha =\vec i)
\end{align}
 where $\vec \alpha$ denotes the vector of those card indices for which $\ket{q_p q_a}$ is in state $\ket{000}$, and the $|$ symbol indicates the control condition:
 \begin{align}\label{eq:cond_op}
     C_{\,\vec i}\,\big| (\vec \alpha =\vec i) = \left\{
     \begin{matrix}
         C_{\,\vec i}, & \mbox{if }\,\vec \alpha =\vec i \\
         I, &\mbox{else.} \end{matrix}\right. 
 \end{align}
Note that $\vec \alpha$ is defined for basis states in the computational basis, and \autoref{eq:cond_op} extends by linearity.

To emphasize that this operator encodes player \textcolor{blue}{A} playing the cards, the notation $CP_k^n=U_k^{\textcolor{blue}{A}}$ will be used.
 
\subsubsection{Uncomputing}
After the cards have been played, the corresponding auxiliary qubits have to be updated as well. The simplest approach is to apply a Pauli-X gate to each $q_a$, conditioned on $\ket{q_tq_s}$ being in state $\ket{10}$.

\subsection{Trick taking}
\label{subsec:tricktaking}
After the cards have been played, the \emph{trick-taking} partial order will determine which player takes the trick. The Hasse diagram of this partial order is shown in \autoref{fig:tricktakingpo}. Whenever the subset of played cards has a unique maximal element, the corresponding player takes the trick. Only when there is more than one maximal element does the order in which the cards were played take precedence. 

\subsubsection{Construction of the $TT_k$ gate}\label{sec:tricktaking}
In the following, a quantum gate is constructed that simulates trick taking in the case of a \emph{totally ordered} $k$-element subset of cards in the trick.  A gate is introduced that implements moving cards from the table onto the round winner's stack. Again, the uncontrolled gate will later be extended to a controlled version. 

Here, the fact is used that the card order can be arranged in the encoding such that, for a totally ordered subset, a higher card is always located to the left of the lower cards, i.e., the leftmost card wins the trick.

The uncontrolled version of the $TT_k$ gate is a $(2+1)k$-qubit gate that acts on those parts of the card state
associated with the location of the card, i.e., $\ket{q_sq_t}$. 

Trick taking consists of two distinct steps: First, each of the $k$ cards is transferred from the table to the stack by evolving its location state from $\ket{q_sq_t}=\ket{01}$ to $\ket{q_sq_t}=\ket{11}$. For a single card, this is achieved simply by applying a Pauli-X gate to the stack qubit $q_s$. 

The next step is to identify the winner of the trick and set the player qubits of all associated cards to the winner's player-qubit state. Suppose that the trick consists of cards with indices $i_1<i_2 < \ldots < i_k$. Due to the total order with respect to trick taking, all player qubits $\ket{q_0^{(i_l)}q_1^{(i_l)}}$ for $l\ge 2$ have to be set to the same state as $\ket{q_0^{(i_1)}q_1^{(i_1)}}$. Since the relevant player-qubit states are known, the No-Cloning Theorem can be avoided by the following construction:
 \begin{description}
    \item[$k=2$] In the case of two players, only one qubit $\ket{q_p}$ is necessary for encoding the players. It suffices to apply a Pauli-$X$ gate to $\ket{q_p^{(i_2)}}$ because $\ket{q_p^{(i_1)}}$ and $\ket{q_p^{(i_2)}}$ are known to be initially in orthogonal states. 
     
    \item[$k>2$] In the case of three or more players, auxiliary qubits are necessary, since $k-1$ orthogonal states corresponding to the losing players are all mapped to the same state, namely the player state of the winner.
     
    The three-qubit $\mbox{CC-NOT}(q_{c_1},q_{c_2}; q_t)$ gate, conditioned on $\ket{q_{c_1}q_{c_2}}$ being either in state $\ket{10}$ or $\ket{01}$, maps states of the form $\ket{q_1}\ket{q_2}\ket{q_1}$ to $\ket{q_1}\ket{q_2}\ket{q_2}$ and thus may be used to rewrite
    the player qubits of a lower-valued card. In this case, for each player qubit, an auxiliary qubit is needed that, up to this part of the circuit, is identical to the player qubit.
    \end{description}

For $i_1<i_2$, let $D_{i_1,i_2}$ denote the unitary gate acting on the player qubits of the $i_1\otimes i_2$ card states and transforming the vectors as
\begin{align*}
\ket{\rm player}_{i_1}\ket{\rm player}_{i_2}& \to
\ket{\rm player}_{i_1}\ket{\rm player}_{i_1}.
\end{align*}

Similarly to the $CP_k^n$ gate in \autoref{eq:card_playing_gate}, all possible unitaries $D_{i_1,i_2}$ for a given $k$-card configuration have to be controlled, and the product over all these configurations is then taken, i.e., the construction is conditioned on states in which exactly $k$ of the kets $\ket{q_{s}^{(l)} q_{t}^{(l)}}$ are in state $\ket{01}$. For this purpose, one auxiliary qubit is needed for each stack qubit $q_s$. For $\vec i = (i_1,\ldots,i_k)\in \mathcal{C}_k$ let
\begin{align}
D_{\vec i} &= \prod_{r=2}^k D_{i_1,i_r}( q_{\vec p_{i_1}},q_{\vec a_{i_r}}; q_{\vec p_{i_r}})\cdot \prod_{l=1}^k X(q_s^{(i_l)})
\label{eq:D_unitary}
\end{align}
where the first factor acts on the trick-losing players' qubits and the second factor on all players' stack qubits. Finally, we define
\begin{align}
TT_k &=  \prod_{\vec i \in \mathcal{C}_k }D_{\,\vec i}\,\big| (\vec \alpha =\vec i).
\label{eq:TTk-gate}
\end{align}
where, analogously to the definition of $\rm CP_k^n$, $\vec \alpha$ denotes those card indices for which $\ket{q_s q_t}$ is in state $\ket{01}$. The implementation of $TT_2$ and $TT_3$ will be described in detail in a later section.
For fixed $k$, the notation $U_{TT}$ is used instead of $TT_k$ when the unitary nature of \autoref{eq:TTk-gate} is to be emphasized.

\subsection{Full and partial game}
A game of Skat with a 32-card deck and three players consists of ten rounds of trick taking. Each round is modeled by the unitary round operator
\begin{align}
    R_i = TT_3\cdot U^C_{11-i}\cdot U^B_{11-i}\cdot U^A_{11-i},
\end{align} where, for the sake of simplicity, the updating of the auxiliary qubits has been omitted. The game evolves by consecutively applying the round operators $R_{10}\cdot\ldots \cdot R_1$. Adaptations for fewer cards, players, and rounds are straightforward.

Often, additional information about a particular game is available, e.g., the starting hand of player $A$ may be known. In this case, a different initial state $\ket{\Phi_{\rm ini}}=U_{\rm ini}\ket{0\ldots 0}$ has to be prepared. Slightly abusing notation, let $C_{i_j}$ denote the operator that plays the $j^{\rm th}$ card from player A's hand. In this way, a different final state is obtained for each $j$:
\begin{align}
    \ket{\Phi_{\rm fin}^j} = R_{10}\cdot\ldots \cdot R_2\cdot \tilde{R}_1^j
\end{align}
where, ceteris paribus, $\tilde{R}_1^j= TT_3\cdot U^C_{10}\cdot U^B_{10}\cdot C_{i_j}$ models the game in which player A plays the $j^{\rm th}$ card from the hand. Again, adaptations for information available in different rounds, etc., should be straightforward to implement.

\begin{figure}[t]
\newcommand{\cards}{A/1,10/2, K/3,Q/4,9/5,8/6,7/7}
\centering
\begin{tikzpicture}[scale=1, transform shape]

   % Upper chain of 8 elements
    \foreach \val/\i in \cards {
        \ifnum\i=1
        \node[draw, circle, fill=white, minimum size =20pt,inner sep = 1pt] 
        (u\i) at (0, 7){$\spadesuit$\val};
        \fi 
        
        \ifnum\i>1
        \node[draw, circle, fill=white, minimum size =20pt,inner sep = 1pt] (u\i) at (0, 8-\i) {$\spadesuit$\val};
            \pgfmathtruncatemacro{\j}{\i-1}
            \draw (u\j) -- (u\i);
        \fi
    }
   %Jacks on top
     \foreach \suit/\i in {\clubsuit/4, \spadesuit/3, \heartsuit/2, \diamondsuit/1} {
        \node[draw, circle, fill=white, minimum size =20pt, inner sep = 1pt] 
        (j\i) at (0, 7+\i){$\suit$J};
     }
     \draw (j2) -- (j1);
     \draw (j3) -- (j2);
     \draw (j4) -- (j3);
     \draw (u1) -- (j1); 
     
    %Chains
    \foreach \val/\i in \cards {
    \foreach \k in {1,2,3} {
        \node[draw, circle, fill=white, minimum size =20pt] (\k l\i) at (-6+3*\k, 1-\i) {};
        \ifnum \k = 1 \node at  (\k l\i) {$\heartsuit$\val}; \fi
        \ifnum \k = 2 \node at  (\k l\i) {$\diamondsuit$\val}; \fi
        \ifnum \k = 3 \node at  (\k l\i) {$\clubsuit$\val}; \fi
        \ifnum\i>1
            \pgfmathtruncatemacro{\j}{\i-1}
            \draw (\k l\j) -- (\k l\i);
        \fi
    }
    }
    %Connections between lower chains
    % \foreach \i in {2,...,7} {
    %    \pgfmathtruncatemacro{\j}{\i-1}
    %    \draw (1l\i) -- ( 2l\j);
    %    \draw (2l\j) -- ( 3l\i);
    %    \draw (2l\i) -- (1l\j);
    %    \draw (2l\i) -- (3l\j);
    %}
    
    %Connections between upper chain and lower chains
    \draw (u7) -- (1l1);
    \draw (u7) -- (2l1);
    \draw (u7) -- (3l1);

    %Example 1
    \foreach \val/\i in {A/1,10/2, K/3,Q/4} {
    \node[draw, circle, fill=blue, minimum size =20pt] at  (3l\i) {$\clubsuit$\val};}
    %    }
    %Example 2
    \node[draw, circle, fill=pink, minimum size =20pt] at  (j4) {$\clubsuit$J};
    \node[draw, circle, fill=pink, minimum size =20pt] at  (j3) {$\heartsuit$J};
    \node[draw, circle, fill=pink, minimum size =20pt] at  (j2) {$\spadesuit$J};
    \node[draw, circle, fill=pink, minimum size =20pt] at  (u7) {$\spadesuit$7};
    \foreach \val/ \i in {A/1,10/2, Q/4,8/6,7/7}{
    \node[draw, circle, fill=pink, minimum size =20pt] at  (1l\i) {$\heartsuit$\val};
    }

\end{tikzpicture}
\caption{Trick-taking partial order for a spades game. Totally ordered subsets for Example 1 (\textcolor{blue}{blue}) and Example 2 (\textcolor{pink}{pink}).}
\label{fig:tricktakingpo}
\end{figure}

\subsection{Evaluation Measurement}

For this section, some basic game rules are needed, such as the values of the cards and the criteria for winning a game. The following table shows the value function $\mathcal{V}$ for each card. Note that the suit does not affect the value.

\begin{table}[h]
\caption{Card values. There are 30 points available for each suit, leading to a total of 120 points in the game. The declarer needs more than half of this total, i.e., 61 points, to win.}
\begin{tabular}{|l|l|l|l|l|l|l|l|l|}
\hline
$\rm Card$ & 7 & 8 & 9 & 10 & J & Q & K & A \\ \hline
$\mathcal{V}$ & 0 & 0 & 0 & 10 & 2 & 3 & 4 & 11 \\ \hline
\end{tabular}

\end{table}

For demonstration purposes, every game is treated as a \textit{spades} game. This is a simplified representation, but it raises no conceptual issues, since other trump-suit choices by the declarer can be related to this problem by introducing simple swap gates. This might increase the circuit depth slightly, but it does not affect the feasibility of the approach.

\subsubsection{Score Operator}
To determine who has won the game, an an observable $\mathcal{S}_A$ (the score operator) is defined that returns the point values of the cards in player A's stack. For each card, let $P_A^i$ denote the projector onto the states in which the player qubit(s) of the $i$th card are in a state indicating that this card is in the stack of player A. Thus,
\begin{align}\label{eq:score_operator}
    \mathcal{S}_A = \sum_{i=1}^{32} \mathcal{V}(i)\cdot P_A^{(i)},
\end{align}
where $\mathcal{V}(i)$ is the value of card $i$. The expectation value $\bra{\Phi_{\rm fin}} \mathcal{S}_A \ket{\Phi_{\rm fin}}$ corresponds to player A's expected score.

A slightly different and probably more practical approach is to ask not for the expectation value of the final score, but for the percentage $p_\text{win}$ of favorable outcomes. This amounts to describing, in a feasible way, the subspace of states considered favorable. Once the projection $P_{\rm fav}$ onto this subspace is available, the overlap $\bra{\Phi_{\rm fin}} P_{\rm fav} \ket{\Phi_{\rm fin}}$ equals the probability of a favorable outcome, i.e., the probability that player A wins the game.

\subsubsection{Quantum Counting}

The quantity $\bra{\Phi_{\rm fin}} P_{\rm fav} \ket{\Phi_{\rm fin}}$ can be obtained from the following observation. If
\begin{align*}
    \ket{\Phi_{\rm fin}} = \sum_{i\in I_{\rm fav}}c_i \ket{\phi_i}+ \sum_{j\in J}c_j \ket{\phi_j}
\end{align*}
is the orthogonal decomposition of $\ket{\Phi_{\rm fin}}$ according to $\mathcal{H} = \mathcal{H}_{\rm fav}\oplus\mathcal{H}_{\rm fav}^{\perp}$, the expectation value can be obtained by identifying how many indices $i\in I_{\rm fav}$ lead to non-vanishing amplitudes $c_i$. 

This is the setup for quantum counting, a combination of Grover's algorithm and quantum phase estimation (QPE), cf. \cite[pp. 261-263]{nielsen2010quantum}. It can be used to obtain a quantum advantage over classical algorithms. The number of paths $N$ in the subset of winning paths can then be counted via
\begin{align}
    N=N^{\rm tot}\cdot\sin^2 \tfrac{\phi}{2},
\end{align}
where $\phi$ is the result of the QPE. This is particularly relevant when solving search problems with an unknown number of hits. This yields an advantage of order $\mathcal{O}(\sqrt{N})$ compared with the classical $\mathcal{O}(N)$ scaling.

In this case, not only the number of winning paths has to be accounted for, but also the associated probabilities $p_i =|c_i|^2$. Thus, a weighted version of quantum counting is needed, and its implementation remains a topic of ongoing research.

Quantum counting is used to sum the winning paths and thereby obtain a reasonable value for the winning probability.
To obtain a reasonable recommendation for a player's action in the game, it is necessary to fix the first card and evaluate the score of the game to obtain a suitable winning probability. To compare the card qualities, the process has to be repeated for each card allowed by the rules.
This expectation value serves as a proper target function to define the \textit{payoff function}, which ultimately has to be maximized.

\subsubsection{Payoff Function}

To understand the outcome of this calculation and how it defines the game-theoretic solution, a suitable payoff function maximized by the approach has to be introduced. The payoff function $\mathcal P$ of player $x$ after $n$ games is more involved and therefore requires further explanation.

\begin{align}
\label{payoff}
    \mathcal P_n(x)=&\sum_{i=1}^n\left(p_\text{win}(x,i)\cdot\text{Val}(x,i)^\text{won}\right.\\\notag
    &\left.-2\cdot\left[1-p_\text{win}(x,i)\right]\cdot\text{Val}(x,i)^\text{lost}\right).  
\end{align}

Depending on the declarer's choice, each suit chosen as trump by player $x$ leads to a different value $\text{Val}(x,i)$ of the $i$th game. It is given by the length of the uninterrupted sequence of present or missing trump cards (beginning with the highest trump card plus one) and multiplied by a multiplier $\mathcal{V}(i)$ that is determined solely by the declarer's choice for the $i$th game (see Table~\ref{table:basic_value}).

\begin{table}[h]
\caption{Basic value of the suits, depending on the choice of the declarer.}
\begin{tabular}{|l|l|l|l|l|l|}
\hline
$\text{Suit}$ & $\diamondsuit$ & $\heartsuit$ & $\spadesuit$ & $\clubsuit$ & $\text{G}$ \\ \hline
$\mathcal{V}$ & 9 & 10 & 11 & 12 & 24 \\ \hline
\end{tabular}
\label{table:basic_value}
\end{table}

It is also important to mention that the value of a \textit{lost} game is \textit{negative} and \textit{doubled}. Therefore, outperforming the opponents and maximizing the expectation value requires a winning probability substantially above $50\%$ for a single game. These percentages shift drastically under the so-called \textit{Seeger-Fabian-System}, which is widely used in professional and semi-professional tournaments. It adds another 50 points for winning and $-50$ points for losing a game, thereby bringing the threshold much closer to the $50\%$ mark.% (see Fig. \ref{payoff_per_game}).

As seen in \autoref{payoff}, it does not matter whether a game is won by a high margin. The only relevant point is whether the game is \textit{won}. Therefore, the goal is to maximize the percentage of games won by player $x$ by finding the best order in which to play the cards. This order is the one that beats the highest number of possible distributions, regardless of the margin.

To check whether an equilibrium state of the game has been found, the payoff function has to be investigated in terms of stability. There should be at least one \textit{allowed} permutation of played cards that leads to a situation in which every player is satisfied with the respective choices according to the information available at that point in the game.

\section{Examples}
The following examples of increasing complexity are analyzed\footnote{The source code for these examples is freely available upon request.}:
\begin{enumerate}
    \item The simplest example is a generic two-player game with the following deck:
\begin{equation}\label{eq:cards_toyexample_four}
        \clubsuit A \hspace{10pt} \clubsuit 10 \hspace{10pt} \clubsuit K \hspace{10pt} \clubsuit Q.
\end{equation}
While the particular selection of cards does not matter much in this setting, it is important to note that the selected cards form a \emph{totally} ordered subset with respect to the trick-taking relation. This first example already helps to represent the superposition of all possible distributions, playing a card, taking the trick, and counting the points at the end of the game. 

\item The second example is a game with three players and six cards. Intuitively, it is not obvious which of the two cards in forehand's hand is better to play.  Following suit will become relevant, and another qubit is needed to store which player took the first trick.

\item The final case is the full game with 32 cards and three players. Since this is beyond the current capabilities of quantum computers, the focus is on a heuristic analysis.
\end{enumerate}

\subsection{Four-Card Toy Example}\label{sec:example_1}
In this section, the quantum circuit corresponding to Example 1, a card game with four cards and two players, is explicitly constructed. 
\subsubsection{Encoding}
The card-state encoding of \autoref{tab:encoding_four_cards} is chosen.
\begin{table}[t]
\caption{Encoding the card state $\ket{j_0j_1 j_2 j_3}$ in the four-card toy example. Here $n_c=3+1$.}
\begin{minipage}{0.35\textwidth}
\centering
\begin{tabular}{|r|c|c|}\hline
        positions & encoded state  \\ \hline
         $j_0$ & player A/player B  \\
         $j_1j_2$ & card location \\
         $j_3$ & auxiliary\\ \hline
\end{tabular}
\end{minipage}
\hfill
\begin{minipage}{0.125\textwidth}
    \centering
\begin{tabular}{|c|c|}\hline
         $j_1j_2$ & loc. \\ \hline
         00 & hand\\ \hline
         10 & table\\ \hline
         11 & stack\\ \hline
\end{tabular}
\end{minipage}

\label{tab:encoding_four_cards}
\end{table}
The cards are arranged according to the trick-taking order. This order is given in \autoref{eq:cards_toyexample_four} and the corresponding values for the scoring are $\mathcal{V}(1) = 11, \mathcal{V}(2) = 10, \mathcal{V}(3) = 4$ and $\mathcal{V}(4) =3$ (cf. \autoref{fig:tricktakingpo} and \autoref{table:basic_value}). The state vector is an element of a $2^{4\cdot 4}$-dimensional Hilbert space and takes the form
\begin{align}
    \ket{\Phi} &= \ket{\phi_{ \rm card\,1}}
    \ket{\phi_{ \rm card\,2}}\ket{\phi_{ \rm card\,3}}\ket{\phi_{ \rm card\,4}}.
\end{align}

For technical reasons, the four auxiliary bits are swapped to the end of the state vector: 
\begin{align}
    \ket{\phi} &=  \ket{\underbrace{q_0 q_1 q_2}_{\rm card \,1} \ldots \ldots\ldots q_{9}q_{10}q_{11}\underbrace{q_{13}q_{14}q_{15}q_{16}}_{\rm auxiliaries}}\\\notag
    &=\ket{\rm card\, 1}\ket{\rm card\, 2}\ket{\rm card\, 3}\ket{\rm card \,4}\ket{\rm auxiliaries}.
\end{align}
The qubit $q_{12}$ is used as an external auxiliary for the implementation of the $SP_{1,2}$ gate and is therefore omitted. 
Consider the following example: The state for a card on the table is $\ket{\textcolor{blue}{0} 10}$ (played by player \textcolor{blue}{A}) or $\ket{\textcolor{brown}{1} 10}$ (played by player \textcolor{brown}{B}). The four card states are encoded using a total of twelve qubits. For example, the qubit state (omitting the auxiliary qubits) $\ket{\textcolor{blue}{0}00\textcolor{blue}{0}00\textcolor{brown}{1}00\textcolor{brown}{1}00}$ decodes to cards 1 and 2 being in the hand of player \textcolor{blue}{A} while cards 3 and 4 are in the hand of player \textcolor{brown}{B}.
%A card in the hand of player \textcolor{blue}{A} has card state $\ket{101\textcolor{blue}{0}}$, while a card in the hand of player \textcolor{brown}{B}  has card state $\ket{101\textcolor{brown}{1} }$.  

\subsubsection{Evolution of the game}
The game consists of different phases, which are summarized in \autoref{tab:phases_four_card}. In each phase, the state vector evolves by applying the corresponding unitary operator, finally yielding
\begin{equation}\label{eq:phases_toyexample}
    \ket{\Phi_{\rm fin}} =   \prod_{i=1}^2\left( U_{\rm TT}\,U^B_i\,U^A_i\right) \,  U_{\rm ini }\,\ket{0\ldots 0}.
\end{equation}  Measurement of the score operator $\mathcal{S}_A$ yields the expected final score of player A $\bra{\Phi_{\rm fin}} \mathcal{S}_A \ket{\Phi_{\rm fin}}$ and player A's probability of winning is given by $\bra{\Phi_{\rm fin}}P_{\rm fav}\ket{\Phi_{\rm fin}}$. In this example, the favorable subspace $P_{\rm fav}(\mathcal{H})$ turns out to be three-dimensional. Each factor of \autoref{eq:phases_toyexample} is now reviewed in detail:

\begin{table}[h]
\caption{Phases of the game and their corresponding unitary transformations.}
\begin{tabular}{|l|l|c|}\hline
phase & content & unitary\\
\hline
Round 0& initial superposition  & $U_{\rm ini}$\\ \hline
Round 1& Player A plays a card & $U^A_2$ \\ \hline
&Player B plays a card & $U^B_2$ \\ \hline
&1~st trick taking& TT${_2}$ \\ \hline
Round 2& Player A plays a card & $U^A_1$ \\ \hline
&Player B plays a card & $U^B_1$ \\ \hline
&2nd trick taking& TT${_2}$ \\ \hline
Round 3& Measurement & - \\ \hline
\end{tabular}
\label{tab:phases_four_card}
\end{table}

\subsubsection{$U_{\rm ini}$: Construction of the initial superposition.}
\begin{figure}[h]
    \centering
    \includegraphics[width=8cm]{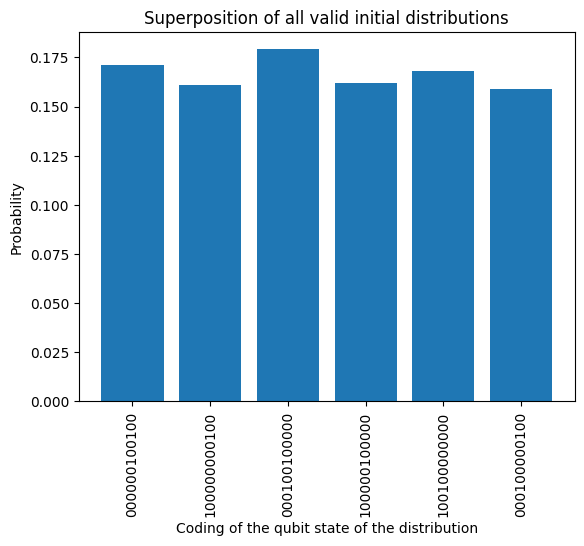}
    \caption{Measurement of $\Phi^{(0)}$ with \numprint{1000} shots.}
    \label{fig:initial}
\end{figure}

First, all initial distributions have to be constructed. Here, one player receives two cards from the four-card deck, and the opponent receives the remaining two cards. There are ${4\choose 2} = 6$ ways to do so.  The superposition of these six basis states is constructed iteratively according to the algorithm proposed in~\cite{superposition}:
\begin{align}
    \ket{\Phi^{(0)}} = \frac{1}{\sqrt{6}}\sum_{i=1}^6 \ket{\phi_i} = U_{\rm ini} \ket{0\ldots 0}.
\end{align}
This results in a superposition with equal probabilities of $\tfrac{1}{6}\approx 0.167$ as shown in \autoref{fig:initial}. 

\subsubsection{$U_2^B U_2^A$: Playing the first card}
\begin{figure}[t]
    \centering
    \includegraphics[width=8cm]{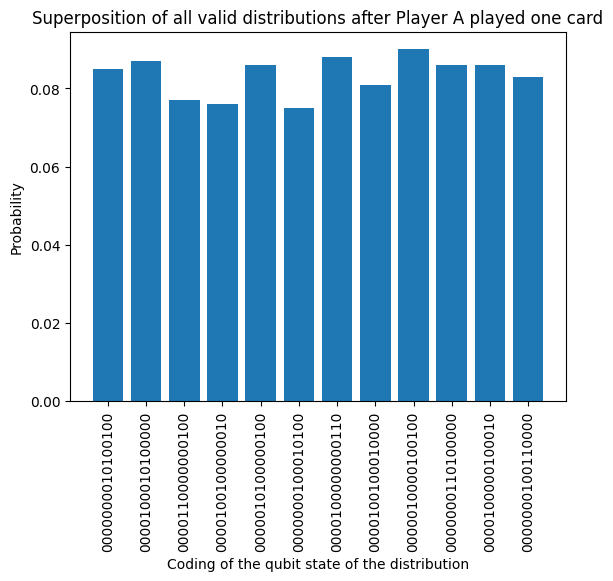}
    \caption{Superposition of the twelve possible card distributions after player A has played the first card using \numprint{1000} shots.}
    \label{fig:firstcard}
\end{figure}

In round 1, player $A$ has two cards in hand. The card-index set is $\{1,2,3,4\}$, so the index set in \autoref{eq:cp_action}
is $\mathcal{C}_2 = \{(i,j): 1\le i< j\le 4\}$. With the chosen encoding, playing a card transforms the card state
\begin{equation}
    \ket{q_0 00} \to \ket{q_0 10},
\end{equation}
which amounts to flipping the middle qubit, i.e., $q_1,q_4,q_7$, or $q_{10}$, depending on the card.
Each factor $CP_2^4$ in \autoref{eq:card_playing_gate}
therefore acts on two qubits via $X_lX_mSP_{l,m}$ for $i,j\in\{1,4,7,10\}$. Note that a version of the circuit has to be applied that is controlled by the condition that exactly two of the player qubits $\{q_0,q_3,q_6,q_9\}$ are in state $\ket{0}$. No conditioning on the table qubits or their auxiliary qubits is needed, because in the first round they are all in state $\ket{0}$. After all factors of the corresponding unitary $U^A_2$ have been applied, the state is an equally weighted superposition of twelve basis states with probabilities of $\tfrac{1}{12}\approx 0.083$, as shown in \autoref{fig:firstcard}. 

The same circuit implements the unitary $U_2^B$ for player B, with the only change being that the condition is now on two player qubits being in state $\ket{1}$. 

\begin{figure}[t]
    \centering
    \includegraphics[width=8cm]{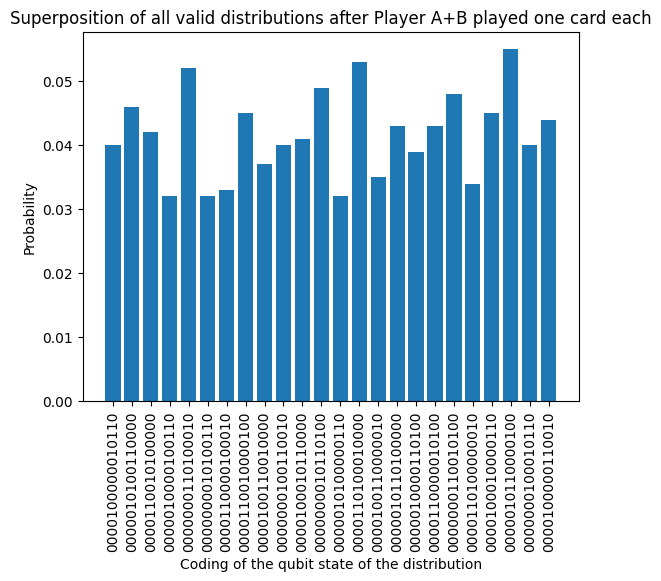}
    \caption{Superposition of the 24 possible card distributions after player B has played the first card using \numprint{1000} shots.}
    \label{fig:secondcard}
\end{figure}

The result is an equally weighted superposition of 24 basis states with probabilities of $\tfrac{1}{24}\approx 0.042$, as shown in \autoref{fig:secondcard}.

\subsubsection{Preparing the auxiliaries}
In this toy example, only one auxiliary qubit per card is needed, which keeps track of whether a card is still in play, i.e., the stack qubits. For simplicity, these auxiliary qubits are updated only at this point. Conditioning can be performed on the stack qubits $q_2,q_5,q_8,q_{11}$ being in state $\ket{1}$, followed by a series of $\cnot$ gates:
\begin{align}
    \prod_{i=1}^4\cnot_{3i-1, 12+i}.
\end{align}

\subsubsection{${\rm TT}_2$: First trick taking}
The two cards on the table can have card indices $1\le i<j\le 4$. The operator 
\begin{align}
D_{i,j} = X_{3(j-1)} 
\end{align}
flips the player qubit of the card with index $j$. This operator is applied only to states in which the cards with indices $i$ and $j$ belong to different players, resulting in a state in which both cards belong to the same player.
The unconditioned unitary from \autoref{eq:D_unitary} is given by
\begin{align}
    X_{3j-3} \cdot X_{3j-1}\cdot X_{3i-1},
\end{align}
e.g., the unitary $D_{1,2}=X_3 X_5 X_2$ maps the state $\ket{\rm card\,1}\ket{\rm card\,2}=\ket{010}\ket{110}$ to the state $\ket{011}\ket{011}$ in which cards 1 and 2 both end up in the stack of player A. The fully conditioned version iterates over all six index configurations $i<j$ in which exactly the qubits $\ket{q_{3i-2}q_{12+i}}$ and $\ket{q_{3j-2}q_{12+j}}$ are in state $\ket{10}$.

\begin{figure}[t]
    \centering
    \includegraphics[width=8cm]{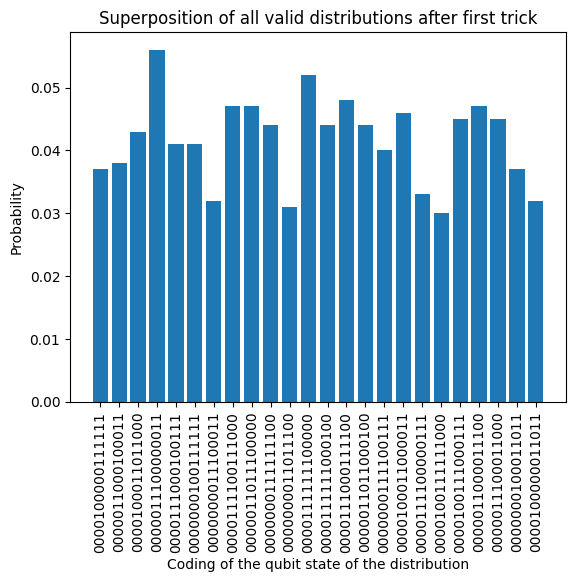}
    \caption{Superposition of the 24 possible card distributions after the first trick using \numprint{1000} shots.}
    \label{fig:firsttrick}
\end{figure}

The resulting state is again an equal superposition of 24 basis states, although the basis states differ from those before the application of $U_{TT}$, cf. \autoref{fig:firsttrick}.

\subsubsection{$U_1^BU_1^A$: Playing the second card}

When playing the last remaining card, there is no choice left for the players. Thus, the unitaries $U_1^{A/B}$ simply flip the qubits  $q_1,q_4,q_7$ or $q_{10}$ subject to the condition that the corresponding auxiliary qubit is in state $\ket{0}$.

\subsubsection{${\rm TT}_2$: 2nd trick taking}
The same unitary as in the first trick taking is applied. All card states are now of the form $\ket{q_p 11}$. The following outcomes are possible:
\begin{itemize}
\item all cards in the stack of player A (probability $\tfrac{1}4$)
\item all cards in the stack of player B (probability $\tfrac{1}4$)
\item two cards in each of the two stacks (probability $\tfrac{1}{12}$ for each of the six possible combinations)
\end{itemize}

\begin{figure}[t]
    \centering
    \includegraphics[width=8cm]{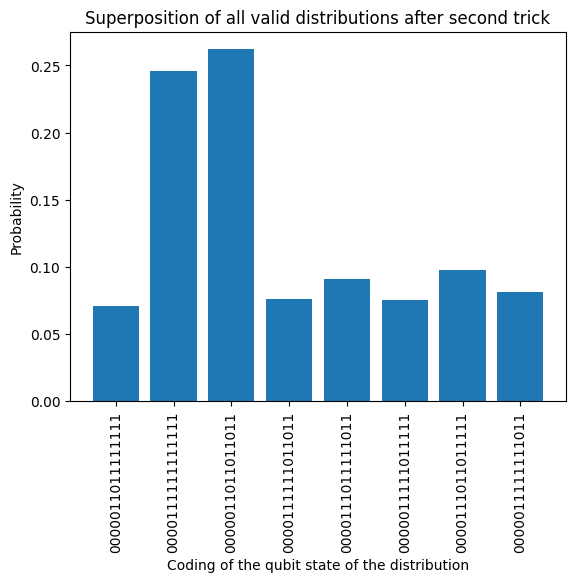}
    \caption{Superposition of the eight possible card distributions after the second trick using \numprint{1000} shots.}
    \label{fig:secondtrick}
\end{figure}

This results in a superposition of eight states, as depicted in Fig.~\ref{fig:secondtrick}.

\subsubsection{Measurement: Scoring}
Finally, the score operator $\mathcal{S}_A$ from \autoref{eq:score_operator} can be applied, returning  the expected point values of the cards in player A's stack. In this particular example, the subspace of favorable outcomes can also be constructed explicitly: player A wins if either $\{\clubsuit A,\clubsuit 10, \clubsuit K,\clubsuit Q\}$, $\{\clubsuit A,\clubsuit 10\}$ or $\{\clubsuit A,\clubsuit K\}$ is in player A's stack. Each situation corresponds to a particular basis state\footnote{Here: $\ket{011,011,011,011},\ket{011,011,111,111},\ket{011,111,011,111}$.} Thus, the subspace is three-dimensional and $p_{\rm win}= \tfrac{5}{12}$.

%%%%%%%%%%%%%%%% DIAGRAM %%%%%%%%%%%%%%%%%%%%%%%%%%%

\subsection{Six-Card Example}\label{sec:example_6}
As a second step, the player count is increased to three, according to the actual game rules. A typical end-game situation is modeled from the perspective of the solo player A in forehand. Player A has acquired 40 points after eight tricks, and has the following cards in hand: 

\begin{equation}\label{eq:cards_six_playerA}
        \clubsuit A \hspace{10pt} \clubsuit K .
\end{equation}

The remaining four cards in the hands of players B and C are:
\begin{equation}\label{eq:cards_six_unknown}
        \clubsuit 10 \hspace{10pt} \clubsuit 9 \hspace{10pt} \clubsuit 8 \hspace{10pt} \heartsuit 10.
\end{equation}

Player A does not know how the four cards are distributed between players B and C and all distributions are assumed to be equally likely. Neither clubs nor hearts are trump.

\begin{table}[h]
    \centering
    \caption{Encoding the card state $\ket{q_{j_0}q_{j_1}q_{j_2}}$}
\begin{tabular}{|r|c|c|}\hline
        positions & encoded state  \\ \hline
         $j_2j_1$ &  fore-, middle-, rearhand, stack\\
         $j_0$ & hand/table or which stack \\
         %$j_5$ & auxiliary\\ 
         \hline
\end{tabular}

\begin{tabular}{|r|c|c|}\hline
        values & position of card  \\ \hline
         000 &  hand player A (solo)\\
         100 &  hand player B\\
         010 &  hand player C\\
         110 &  team stack\\
         001 &  table player A\\
         101 &  table player B\\
         011 &  table player C\\
         111 &  solo stack\\

         \hline
\end{tabular}\label{tab:encoding6cards}
\end{table}

Intuitively, playing the ace may be an option for player A. However, it turns out that playing the king gives better winning chances.

The encoding of the card states used here is slightly more complicated than in the four-card example. To fit the encoding into three qubits per card, the stacks of players B and C are not separated, since they play together. In this way, the separation between the location qubits and the player qubits is lost in the case of the stack. In exchange, three qubits per card are sufficient, i.e., 18 qubits plus auxiliary qubits, which can be executed on a laptop simulator. Otherwise, two qubits for the player and two qubits for the state would result in 24 plus auxiliary qubits, which may be beyond the limits of a regular laptop computer.

\begin{figure}[t]
    \centering
    \includegraphics[width=8cm]{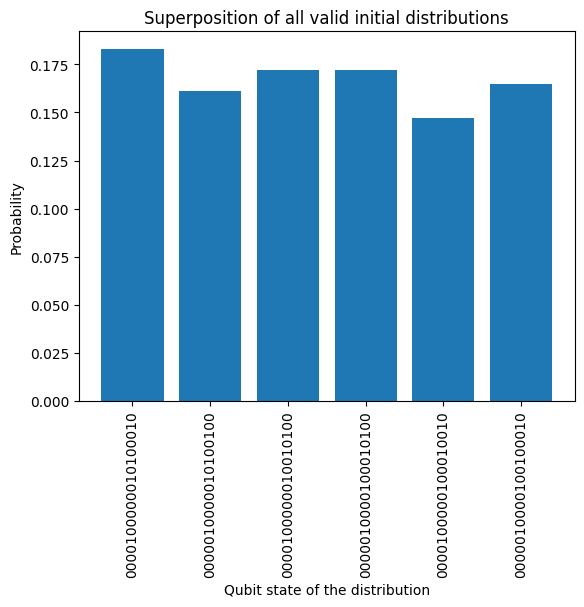}
    \caption{Superposition of the six possible initial distributions using \numprint{1000} shots.}
    \label{fig:initialsix}
\end{figure}

The initial distribution has again ${4\choose 2} = 6$ possibilities. The state 000 100 000 100 010 010 means that the ace and king of clubs are in the hand of player A ("000"), while the 10 and 9 of clubs are in the hand of player B ("100"). The remaining two cards are in the hand of player C ("010"). A superposition of the six possible encoded states is therefore generated, as shown in \autoref{fig:initialsix}.

\begin{figure}[t]
    \centering
    \includegraphics[width=8cm]{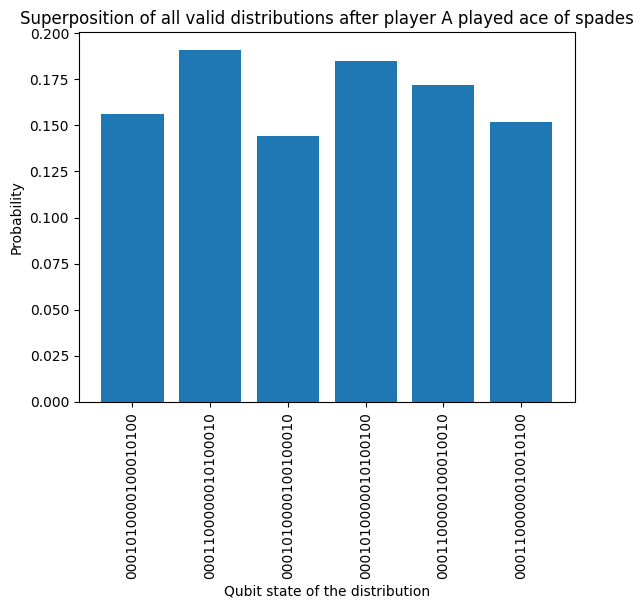}
    \caption{Superposition of the possible card distributions after player A has played the ace of clubs, using \numprint{1000} shots.}
    \label{fig:firstcardsix}
\end{figure}

Playing the ace as player A's first card in forehand results in an equal distribution over six possible states; see \autoref{fig:firstcardsix}. This simply flips the first three qubits from 000 to 001.

\begin{figure}[t]
    \centering
    \includegraphics[width=8cm]{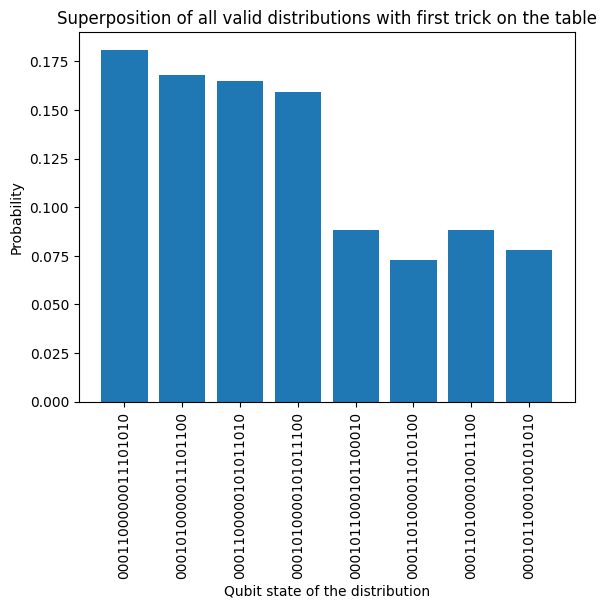}
    \caption{Superposition of the possible card distributions after the first trick on the table, using \numprint{1000} shots.}
    \label{fig:firsttrickontablesix}
\end{figure}

The player holding the 10 of hearts must follow suit and can only play the clubs card. The other player has two options, so there are eight possible distributions for the first trick on the table; see \autoref{fig:firsttrickontablesix}. Again, controlled operations are used to check whether a card is in the hand of player B or C.

Taking the first trick converts all cards on the table to 111 if player A played the ace, or to 110 if player A played the king. 

In the latter case, the player with the 10 of clubs will always play it. However, an additional qubit must now be used to store whether player B or C played the 10 of clubs, as this is necessary for trick 2. Thus, besides the qubits indicated in \autoref{tab:encoding6cards}, one additional qubit is needed for taking the first trick (0: player B, 1: player C). If player A takes the first trick, it is irrelevant who takes the second trick.

\begin{figure}[t]
    \centering
    \includegraphics[width=8cm]{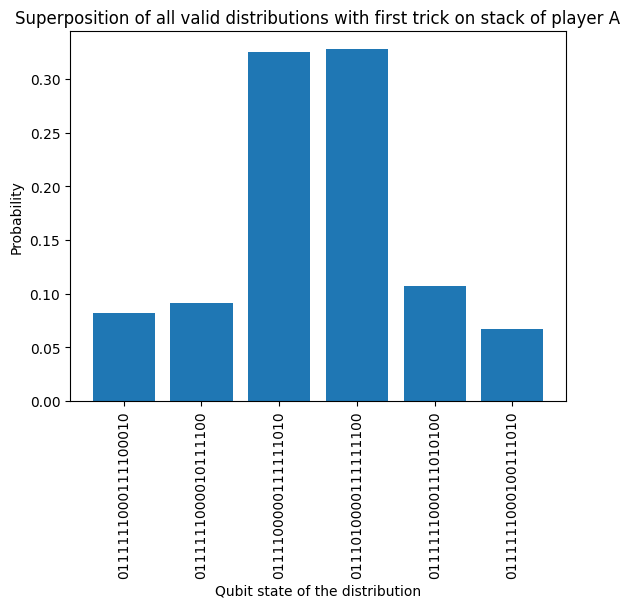}
    \caption{Superposition of the possible card distributions after the first trick in the stack of player A using \numprint{1000} shots.}
    \label{fig:firsttrickonstacksix}
\end{figure}

The distribution after the first trick, with player A having played the ace, is shown in \autoref{fig:firsttrickonstacksix}. 

\begin{figure}[t]
    \centering
    \includegraphics[width=8cm]{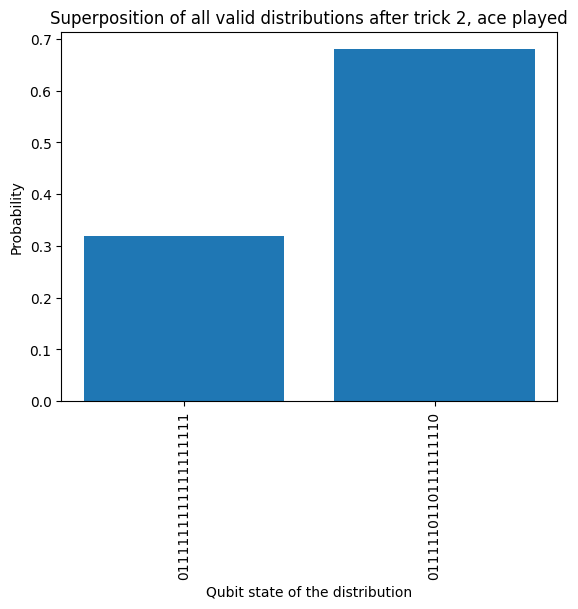}
    \caption{Superposition of the two possible card distributions after the second trick with the ace played initially by player A, using \numprint{1000} shots.}
    \label{fig:secondtricksixace}
\end{figure}

\begin{figure}[t]
    \centering
    \includegraphics[width=8cm]{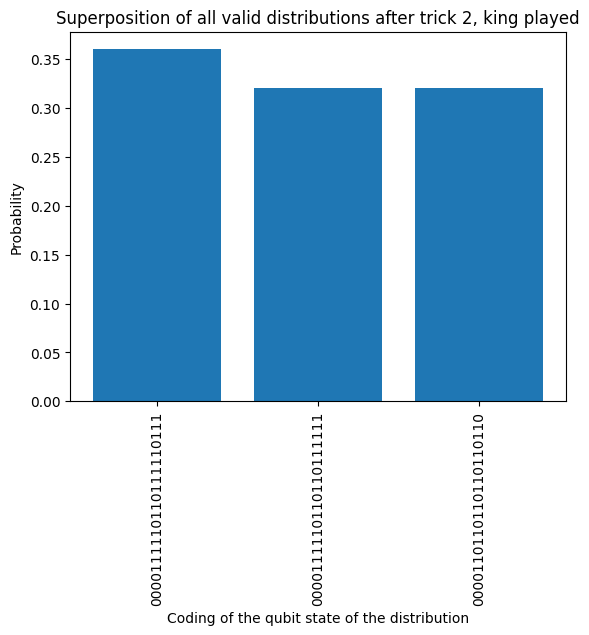}
    \caption{Superposition of the three possible card distributions after the second trick with the king played initially by player A, using \numprint{1000} shots.}
    \label{fig:secondtricksixking}
\end{figure}

The second trick can then also be played and placed in the corresponding stack. The outcome depends on the first card played by player A. If player A played the ace, the final distribution after the second trick is shown in \autoref{fig:secondtricksixace} with two possible outcomes. Only the first outcome (111 111 111 111 111 111) corresponds to a win for player A, with probability $1/3$. If player A initially plays the king, the game results in the distribution shown in \autoref{fig:secondtricksixking}. There are three equally likely possible outcomes, two of which indicate a win for player A. Only 110 110 110 110 110 110, corresponding to all cards being with players B and C, is a losing case for player A. Therefore, player A wins with probability $2/3$.

\begin{table}[h]
\caption{Phases of the game and their corresponding unitary transformations.}
\begin{tabular}{|l|l|c|}\hline
phase & content & unitary\\
\hline
Round 0& initial superposition  & $U_{\rm ini}$\\ \hline
Round 1& Player A plays a card & $U^A_2$ \\ \hline
&Player B plays a card & $U^B_2$ \\ \hline
&Player C plays a card & $U^C_2$ \\ \hline
&1~st trick taking& TT${_3}$ \\ \hline
Round 2& Players play their cards& $U^A_1$ $U^B_1$ $U^C_1$\\ \hline
&2nd trick taking& TT${_3}$ \\ \hline
Round 3& Measurement & - \\ \hline
\end{tabular}
\label{tab:phases_six_card}
\end{table}

%%%%%%%%%%%%%%%% DIAGRAM %%%%%%%%%%%%%%%%%%%%%%%%%%%

\section{Quantum Feasibility of the Full Game}

The quantum nature of the state leads to a probability distribution. Thus, each state is measured with a certain probability. To reduce statistical errors, the quantum circuit is executed several times on the same state.

As described above, all three players receive ten cards, with two additional cards accessible through a bidding process. Considering only the raw numbers, this gives
\begin{align*}
\mathcal{N}=\binom{32}{10}\binom{22}{10}\binom{12}{10} = \frac{32!}{10!10!10!2!} \\ = \numprint{2753294408504640} \approx 2.753\cdot 10^{15}
\end{align*}
different deals. Thus, there are roughly 2.75 quadrillion possibilities for the cards to be distributed among the three players, neglecting the three possible positions of the issuer.
Following the well-known
\emph{birthday paradox}, the probability $p$ that at least one deal is repeated among $k$ deals is
$p = 1 - (\prod_{i=0}^{k-1} (n-i)/n^k)$. Fixing
$p \ge 50\%$ yields $k \ge 40$ million games, making a repeated deal highly unlikely. 

In an empirical analysis, hundreds of millions of Skat games were played with three AIs, 
with an average CPU time of approx.\ 3~s per game on a one-core contemporary PC, 
which corresponds to roughly 1~s of analysis time per player per game. Even if this computation time is reduced to 0.1~s by using less analysis and more computational power, the scaling of the problem still causes a computationally infeasible tree search. Computing the card proposals for 
a selected player across all $2.753\cdot 10^{15}$ deals and dividing this number by
$(3600 \times 24 \times 365)$ gives about \numprint{8730639} years.
Search mechanisms for the players employ, e.g., mini-max search to find the
value of an open game, assuming full card knowledge and collaboration between the two opponent players, followed by voting on different samples~\cite{DBLP:conf/ki/Edelkamp24}, search on the partition of the cards into suit factors~\cite{DBLP:conf/ki/Edelkamp20}, 
or a so-called paranoia search, in which opposing cards are sampled in the search tree 
and, e.g., information about failure to follow suit is propagated~\cite{DBLP:conf/cig/Edelkamp21}. These tree search algorithms do not count the exact number of winning branches in the search tree, but propagate the game value
bottom-up. Moreover, search in practice relies on pruning via transposition tables, alpha-beta search, or equivalents such as moving target search. 
  
To approximate the crossover between classical and quantum computation time, a reduced Skat game with $3x+2$ cards is considered, which amounts to 
$\frac{(3x+2)!}{x!x!x!2!}$ initial deals, even though the playing rules of 
such a reduced game would depend on the cards omitted from the deck. 
Assuming 0.1~s per game as 
a constant gives the following approximation for the classical 
computation time. \\

\begin{table}
 \caption{Estimate of the computation time for all paths over all distributions.}
\begin{tabular}{|c|c|c|c|} \hline
Cards & Deals & Time/Game & Total\\ \hline
 $10$ & $\approx 2.753\cdot 10^{15}$  & 0.1~sec & $\numprint{8730639}$ years \\
 $9$ & $\approx 9.251\cdot 10^{13}$ & $<0.1$~sec & $<\numprint{293368}$ years \\
 $8$ & $\approx 3.0763 \cdot 10^{12}$ & $<0.1$~sec & $<\numprint{9755}$ years \\
 $7$ & $\numprint{100965458880}$ & $<0.1$~sec & $<\numprint{320}$ years \\
 $6$ & $\numprint{6518191680}$ & $<0.1$~sec & $<20.67$ years \\
 $5$ & $\numprint{102918816}$ & $<0.1$~sec & $<119$ days \\ 
 $4$ & $\numprint{3153150}$ & $<0.1$~sec & $<3.65$ days \\
 $3$ & $\numprint{92400}$ & $<0.1$~sec & $<2.57$ hours \\ 
 \hline
\end{tabular}
\label{table:total_time}
\end{table}

\smallskip

Table~\ref{table:total_time} gives a rough estimate of the computation time for the entire game of Skat. Therefore, this estimate suggests that it would take up to 8.7 million years to compute all game paths for all possible distributions, assuming one tenth of a second of computation time per game.
\smallskip
Note that these values are very rough estimates of the classical computation time,
as the time per game will also decrease for smaller numbers of cards in the deal. However, in a real AI-player implementation, 
fewer cards typically lead to more analysis, e.g., earlier search  
to find forced wins or the drawing of more samples to counterbalance
the uncertainty in the unknown cards during voting.

\subsection{Partial Knowledge}

Nonetheless, since the problem is to be solved for one player with given cards, that number can be reduced substantially.

As mentioned previously, the basic idea in a game with imperfect information is not to find \textit{the} single way to win a particular game, but to maximize the expectation value of the payoff function. Success requires maintaining a high level of concentration throughout long tournaments in order to make correct decisions more often than the opponents. Interestingly, a decision can be \textit{correct} in expectation while turning out to be wrong in a particular realization. After all, the \textit{solution} to a given game situation is to find the card that leads to a win, and therefore to points, in most of the remaining scenarios.

For a given hand of player A, the problem size for a strategy recommendation is substantially reduced, since the number of possibilities is divided by $\binom{\text{total number of cards}}{\text{cards in player A's hand}}$.\\

\smallskip

Therefore, only distributions that remain possible given the player's own cards are taken into account.

\begin{table}
     \caption{Estimate of the computation time for all paths over distributions that do not include cards of player A.}
\begin{tabular}{|c|c|c|c|}\hline
Cards & Deals & Time/Game & Total\\ \hline
 $10$ & $\numprint{42678636}$  & 0.1~sec & $49.28$ days \\
 $9$ & $\numprint{9237800}$ & $<0.1$~sec & $<10.69$ days \\
 $8$ & $\numprint{1969110}$ &$<0.1$~sec & $<2.28$ days \\
 $7$ & $\numprint{411840}$ & $<0.1$~sec & $<11.45$ hours \\
 $6$ & $\numprint{168168}$ & $<0.1$~sec & $<4.67$ hours \\
 $5$ & $\numprint{16632}$ & $<0.1$~sec & $<27.72$ min \\ 
 $4$ & $\numprint{3150}$ & $<0.1$~sec & $<5.25$ min \\
 $3$ & $\numprint{560}$ & $<0.1$~sec & $<56$ sec \\ 
 \hline
\end{tabular}
\end{table}

For the total number of remaining states $\widetilde{\mathcal{N}}$ this gives
\begin{align}
\mathcal{\widetilde{N}}=\binom{22}{10}\binom{12}{10} = \numprint{42678636}.
\label{eq:reduced_possibilities}
\end{align}
%One has to take into account, that, up to this point, we were only focusing at different distributions of the cards, while the actual interesting part is to get to know what paths of the decision tree lead to which result.
It is worth noting that solvers for given situations and distributions already exist. However, this does not necessarily mean that they reflect the way a card \textit{should} be played in order to maximize the expected payoff.

\subsection{Upper Bound}

Consider three players playing the game, each equipped with an individual quantum computer. If all players act $100\%$ rationally, they should tend toward a Nash equilibrium, consecutively getting better results as the game proceeds, since the information available during the game increases over time. The state in which the entire distribution is known may also occur before step ten, depending on the present distribution. In that case, the superposition of possible states enters the classical limit and can be solved using existing devices. In earlier stages, in which distributions that differ substantially from the actual one are still possible, this may lead to different conclusions, producing suggestions that are not valuable in that particular case but advantageous in the long run.

It has to be mentioned that this approach considers only perfect play by both players. This might be an unrealistic scenario for a real-game approach, since the partner may misinterpret the card. It has to be taken into account that, in a real game, \textit{not all remaining possible distributions are equally likely} because the opponents usually seek attacking opportunities. Therefore, the probabilities change, and the paths have to be weighted according to the cards that appear on the table.

Since the rules of the game do not allow all cards to be played at all times, this is only an upper bound for the number of possibilities and the computational resources required. In fact, for an actual game prediction, the required computational effort is much lower, since the rules significantly reduce the available options.

The upper bound for this scenario is straightforward to derive. If \textit{every} player were allowed to play \textit{every} single card at \textit{all times}, each player would have a total of 
\begin{align}
N_\text{i}=10!
\end{align}
options. This gives the following number of possible paths in the decision tree for a given distribution:
\begin{align}
N_\text{tot}=(10!)^3.
\end{align}
As already seen in \autoref{eq:reduced_possibilities}, the total number of distributions after the player's own cards are known is \numprint{42678636}. This leads to a total of
\begin{align}
\widetilde{N}_\text{tot}=\numprint{2039386920479691374592000000}%\approx 2\cdot 10^{27}
\end{align}
possible games with 22 of the 32 cards unknown.

As mentioned above, this number is an absolute upper bound, since the game rules strongly restrict the paths that a game can take. It can range from an unrestricted card choice to a forced card with no alternatives. Taking all these considerations into account, this large solution space is reduced drastically, depending on the present distribution. 
\subsection{Quantum Optimization}
In principle, each card-play gate $CP_k^n$ can be turned into a parameterized circuit $CP_k^n(c_1,c_2,\ldots,c_k)$, where the $c_i$ are complex amplitudes, replacing the factor $\tfrac{1}{\sqrt{k}}$ in \autoref{eq:cp_action}. Applying a variational quantum eigensolver (VQE) algorithm to this circuit, equipped with a suitable cost function, could lead to a more realistic assignment of weights to particular paths. However, this approach is left for future research.  

\subsection{Experimental Analysis}

For the experimental analysis on one core of an AMD EPYC 7702P processor at 1.977 GHz, an automated Skat engine was employed that has been shown to challenge human playing strength~\cite{DBLP:conf/cg/Edelkamp22}. 
The empirical investigation of the average number of options available to the three players in a Suit game yields $\mu_{\rm Suit} \approx 2.844$ cards. For Grand games, a slightly higher mean branching factor $\mu_{\rm Grand} \approx 2.994$ was obtained.

To obtain this number, $\numprint{1000}$ human suit games were rerun by simulation, using the geometric mean for the number of options at any move within a single game. 
Let $b_{i,j}$ be the branching factor recorded in trick $i \in \{1,\ldots,30\}$ of game $j \in \{1,\ldots,1000\}$. We have
\begin{align}
\mu_{\rm Suit} = \frac{\sum_{j=1}^{1000}\sqrt[30]{ \prod_{i = 1}^{30} b_{i,j}}}{1000} \approx 2.844.
\label{eq:mu_suit}
\end{align}
The analysis took slightly more than 100~min. 
Assuming this to be a good estimate, the number of paths over all possible distributions leads to a total of
\begin{align}
\widetilde{N}^{\text{red}}_\text{tot}\approx \mu_{\rm Suit}^{30}\binom{22}{10}\binom{12}{10}\approx 1.77\cdot 10^{21} \approx 1.5\cdot 2^{70}.
\end{align}

This means that, in a more efficient encoding, the number of qubits imposed by the game might be much smaller than suggested.
Furthermore, the number $N^\text{App}$ of applications of the algorithm required to make an actual suggestion is lower than the maximum number
\begin{align}
N^\text{App}_\text{max}=55=10\cdot \left(10+1\right)/2.
\end{align}
According to Eq.~\eqref{eq:mu_suit}, it is roughly half of this value:
\begin{align}
\overline{N^\text{App}}\approx 10\cdot 2.844 =28.44.
\end{align}

In addition, it is possible to redistribute the probabilities for each card being in the hand of player A, B, or C when the distribution is reevaluated using the obtained information. Every piece of information gained during the game may contribute to eliminating more and more paths from the decision tree until only a single path remains for the final card.

As the individual knowledge of each player increases during play, e.g., when other players do not follow trump or suit, the maximum number of possible games $\binom{22}{10}\binom{12}{10}$ decreases rapidly during a game. It should also be noted that after bidding and game selection, as well as Skat putting, the distribution of cards is no longer uniform.\medskip

\section{Complexity of the algorithm}

To estimate the complexity of the quantum algorithm, the steps of \autoref{tab:phases_four_card} are considered, and the CX gates necessary for their execution are counted.

\subsubsection{Initial superposition}
\begin{figure}[t]
    \centering
    \includegraphics[width=9cm]{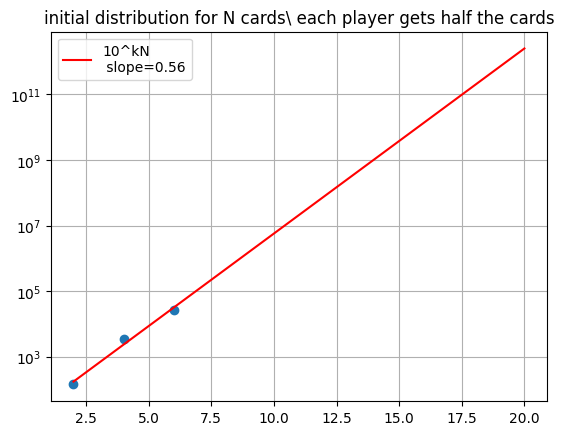}
    \caption{Extrapolation of the complexity of the initial superposition.}
    \label{fig:initialsuperposition}
\end{figure}

The complexity of the algorithm described in~\cite{superposition} depends on the Boolean function chosen. In this case, the cards of player A are known. Players B and C both have $N$ out of $2N$ cards (neglecting the Skat). This has been calculated for $N=1,2,3$ and extrapolated as shown in Fig.~\ref{fig:initialsuperposition}. For $2N = 20$, the complexity is of order $10^{12}$.

\subsubsection{Player A plays a card}
Player A has $N$ cards, so there are $N$ options, with $N=10,9,\ldots,1$ over the ten tricks.

\subsubsection{Player B/C plays a card}
The card-playing mechanism in $U^X_N$ is executed $\binom{2N}{N}$ times. Each execution involves a multi-controlled gate MCX$(N)$ with $N$ control bits. It is assumed that the complexity of such a gate is $12N$ times the complexity of a CX gate.
Assuming that on average there are $\frac{N}{4}$ playable-card options due to the suit-following rule, this yields a quantity of order $\binom{2N}{N} \frac{N}{4} 12N \approx 4^N N$ CX gates

\subsubsection{Trick taking}
The trick-taking function is also executed $\binom{2N}{N}$ times. Each execution involves an MCX gate with $2N$ control qubits, resulting in $24N$ CX gates.
The trick-taking operation itself requires three CX gates, i.e. a negligible quantity. 

\subsubsection{Overall}
For $N=10$, card playing and trick taking lead to a CX count approximately of order $10^7$. Compared with the initial-superposition complexity of $10^{12}$, this is negligible.

The whole circuit has to be executed with \numprint{1000} shots, or even more, to obtain a good approximation. In total, this gives around $10^{16}$ CX gates. Using a state-of-the-art timescale for two-qubit gates on a superconducting QPU of about $\sim 40$~ns, according to~\cite{benchmark_iqm}, this corresponds to approximately twelve years of calculation time. 

However, in the present form, the circuit complexity stems primarily from the preparation of the initial superposition. While examples exist in the literature on how to prepare equal superpositions of $M$ states with only $\mathcal{O}(\log M)$ CX gates, there are certain assumptions on the subspace of superposed states that make the problem tractable, cf. \cite{Shukla2024UniformSuperposition}. Therefore, further research is warranted to determine whether the presented encoding allows such a low-complexity realization, or whether modifying the encoding itself is the more promising approach. As an intermediate step, it seems sensible to investigate the approximability of the initial superposition, either by transforming into a delocalized set of basis states and truncating or by using matrix product states.  It should be noted that the generation of equal superpositions of certain states is currently an active research area. 

With this in mind, the algorithm for ten cards would be in the same order of magnitude as the classical case. For larger setups, advantages can be expected from the quantum algorithm.

\section{Conclusion}

It has been shown that the game information of an imperfect-information game such as \textit{Skat} can be encoded into quantum states, and that the game's initial state can be prepared as a superposition using unitary gates. The uncertainty of the random deal is represented by this initial quantum register.
Trick taking can also be modeled, and the resulting count can be measured and mapped to the subspace of winning distributions. 

The calculations used to evaluate the belief space for incomplete-information card-game play
can therefore be executed efficiently on a quantum computer.  Evaluation of states is possible through a model-counting procedure based on comparison with the winning subspace after measurement of the score. Computational aspects related to reducing the number of paths in the quantum computation have also been discussed.

According to rough estimates, a quantum advantage may arise in the tree search as the game size increases. Since the full game is played with ten cards per player, a quantum advantage may become relevant at this scale in the future.

More generally, iterating over different initial states and game rules for the various game types may yield exploration results that influence bidding, Skat putting, and game-selection strategies.
Quantum computation can therefore be used to recommend a card to play in the game. 

The aims of this paper are more general. Skat was chosen as a prototype of a selected
partially observable planning problem to show the applicability of the approach. There is likely to be 
a wider range of other problems for which the derivations used to exploit quantum computational gains 
will prove helpful. This work is therefore best interpreted as a theoretical 
and pedagogically accessible study of quantum feasibility for solving partially observable problems.

So far, the exposition does not include an implementation on a real quantum computer, but applications of quantum computing to incomplete-information games can be expected in the foreseeable future.  

\subsection{Outlook}

This work should be regarded as a first attempt to tackle the problem using quantum algorithms with a view toward quantum advantage.
Another aspect for future investigation is a suitable payoff function for player $A$. This includes suggestions for suit declaration according to the player's hand. A more complex problem would be suggesting how to put the \textit{Skat}. This increases the tree size considerably since the hand seen by the player is no longer in a defined state but rather in a $\binom{12}{10}$-superposition.\\
In addition, a suitable implementation in a transpiler for a given gate set and topology on a real QPU has to be investigated to obtain a better understanding of the final circuit depths and necessary coherence times. Future work may also discuss collaboration opportunities among players. For classical search, maintaining knowledge vectors and using weighted sampling techniques in the belief space are known to be effective. Moreover, the approach lacks a formal proof of the stability of the suggested solution.

\subsection{Note Added}

It should be emphasized that the main focus was the feasibility in principle of solving Skat, as an example of an imperfect-information card game, with a quantum computer. Many major features of the game and the corresponding game rules themselves were not included, such as additional information from the bidding process, picking up or putting the Skat, raising bids through so-called \emph{hand} games or even tactical situations that require completely different actions by the players in order to win.

\section*{Acknowledgements} 
 
The authors appreciate discussions with M.~Bauer, M.~Knufinke, I.~Seipp, J.~Wehling, S.~Seegerer, A.~Hirschmeier, and W.~T.~Strunz. Special thanks go to H.~Auer for his networking efforts within the Skat community. Stefan Edelkamp's contribution was supported by GA\v{C}R project 24-12046S.

\section*{References}

\bibliographystyle{unsrt}
\bibliography{references}

\end{document}